\documentstyle[11pt,aaspp4]{article}

\begin{document}     

\def\today{\ifcase\month\or
January\or February\or March\or April\or May\or June\or
July\or August\or September\or October\or November\or December\fi
\space\number\day, \number\year}

%\slugcomment{DRAFT -- \today}

\baselineskip 0.185in

\newcommand{\squig}{$\sim$}
\newcommand{\squigleq}{\mbox{$^{<}\mskip-10.5mu_\sim$}}
\newcommand{\squiggeq}{\mbox{$^{>}\mskip-10.5mu_\sim$}}
\newcommand{\squiggeqmm}{\mbox{$^{>}\mskip-10.5mu_\sim$}}
\newcommand{\decsec}[2]{$#1\mbox{$''\mskip-7.6mu.\,$}#2$}
\newcommand{\decsecmm}[2]{#1\mbox{$''\mskip-7.6mu.\,$}#2}
\newcommand{\decdeg}[2]{$#1\mbox{$^\circ\mskip-6.6mu.\,$}#2$}
\newcommand{\decdegmm}[2]{#1\mbox{$^\circ\mskip-6.6mu.\,$}#2}
\newcommand{\decsectim}[2]{$#1\mbox{$^{\rm s}\mskip-6.3mu.\,$}#2$}
\newcommand{\decmin}[2]{$#1\mbox{$'\mskip-5.6mu.$}#2$}
\newcommand{\asecbyasec}[2]{#1$''\times$#2$''$}
\newcommand{\aminbyamin}[2]{#1$'\times$#2$'$}

\title{Infrared Candidates for the Intense Galactic X-ray Source GX\,17+2\,\footnote{\ Based on observations with the NASA/ESA Hubble
Space Telescope, obtained at the Space Telescope Science Institute,
which is operated by the Association of Universities for Research in
Astronomy, Inc., under NASA contract NAS5-26555.}
}
\author{Eric W. Deutsch, Bruce Margon, and Scott F. Anderson}
\affil{Department of Astronomy, 
       University of Washington, Box 351580,
       Seattle, WA 98195-1580\\
       deutsch@astro.washington.edu; margon@astro.washington.edu;
       anderson@astro.washington.edu}

\author{Stefanie Wachter}
\affil{Cerro Tololo Inter-American Observatory, 
       National Optical Astronomy Observatories\,\footnote{\ Operated
         by the Association of Universities for Research in Astronomy,
         Inc., under cooperative agreement with the National Science
         Foundation.},
       Casilla 603, La Serena, Chile\\
       swachter@noao.edu}

\author{and\\ \vskip .1in W. M. Goss}
\affil{National Radio Astronomy Observatory\,\footnote{\ The National
       Radio Astronomy Observatory is a facility of the National Science
       Foundation operated under cooperative agreement by Associated
       Universities, Inc.},
       P.O. Box 0, Socorro, NM 87801\\
       mgoss@aoc.nrao.edu}

\begin{center}
Accepted for publication in The Astrophysical Journal\\
%To appear in volume 493, 1998 February 1\\
{\it received 1999 February 11; accepted 1999 May 10}
\end{center}

%==============================================================================
\clearpage
\begin{abstract}

We present new astrometric solutions and infrared {\it Hubble
Space Telescope} observations of GX\,17+2 (X1813--140),  one of the
brightest X-ray sources on the celestial sphere. Despite 30 years of
intensive study, and the existence of a strong radio counterpart with
a sub-arcsecond position, the object remains optically unidentified.
The observed X-ray characteristics strongly suggest that it is a so-called
``Z-source," the rare but important category that includes Sco~X--1 and
Cyg~X--2. Use of the USNO-A2.0 catalog enables us to measure the position
of optical and infrared objects near the radio source to sub-arcsecond
precision within the International Celestial Reference Frame, for direct
comparison with the radio position, which we also recompute using modern
calibrators.  With high confidence we eliminate the $V$$\sim$$17.5$ star NP~Ser,
often listed as the probable optical counterpart of the X-ray source,
as a candidate.  Our {\it HST} NICMOS observations show two faint
objects within our \decsec{0}{5} radius 90\% confidence error circle.
Even the brighter of the two, Star A, is far fainter than expected
($H\approx19.8$), given multiple estimates of the extinction in this
field and our previous understanding of Z sources, but it becomes the
best candidate for the counterpart of GX\,17+2. The probability of a
chance coincidence of an unrelated faint object on the radio position
is high. However, if the true counterpart is not Star A, it is fainter
still, and our conclusion that the optical counterpart is surprisingly
underluminous is but strengthened.

\end{abstract}


%==============================================================================
\clearpage
\section{INTRODUCTION}

One of the brightest persistent X-ray sources in the Galaxy (and thus in
the entire X-ray sky), GX\,17+2 (X1813--140) has been observed since
the early sounding rocket days (Friedman et al.~1967). The object
is now deemed a classic low-mass X-ray binary, exhibiting variable,
non-thermal radio emission, X-ray bursts, and ``Z-source" characteristics,
including quasi-periodic X-ray oscillations. As the group of Z-sources
contains many of the X-ray brightest objects in the sky, including Sco
X--1 and Cyg X--2, but is numerically small (about a half-dozen), it is
clearly desirable to identify and study optical counterparts of as many
as possible.

GX\,17+2 was in fact ``optically identified" more than a quarter-century
ago (Tarenghi \& Reina 1972) with a $V$$\sim$$17.5$ G~star, now known as
NP~Ser, on the basis of an excellent X-ray position, and subsequently
a sub-arcsecond radio position (Hjellming 1978). There is one problem,
however: the optical ``counterpart" stubbornly refuses to show any
photometric or spectroscopic abnormalities (e.g., Davidsen et al. 1976;
Margon 1978; Cowley, Hutchings, \& Crampton 1988; Bandyopadhyay et
al. 1999), despite the estimate of $L_x/L_{opt}\sim3000$ (Bradt \&
McClintock 1983). Imamura et al. (1987) reported a single, odd optical
spike of three minutes duration in a $10''$ aperture around this object,
an observation to our knowledge not replicated or confirmed in the
last decade. Naylor et al. (1991) reported possible IR variability,
and colors inconsistent with a single, normal star.

At this low galactic longitude and latitude ($l=16.4, b=+1.3$)
projected near the galactic center, GX\,17+2 is a clear case where
the ``optical identification" could be a chance superposition of an
unrelated object. Indeed, Naylor et al. suggest NP~Ser may be such
a superposition on the X-ray source, based on incompatible absorption
inferred from the optical and X-ray objects. Penninx et al. (1988) opine
that there is no plausible optical counterpart.  On the other hand, some
very current reviews (e.g., van Paradijs 1995) still tabulate NP~Ser as
the identification.
 
Using high angular resolution imagery of the {\it Hubble Space Telescope
(HST)} WFPC2, Deutsch et al. (1996) find that in the visible, NP~Ser
appears to be a single object, with no significant deviations from
the typical WFPC2 point-spread function (PSF) of \decsec{0}{074} FWHM.
By subtracting a model PSF, they set upper limits on a possible second
nearby object which might be the real optical counterpart, with $R>23.5$
at angular separation larger than \decsec{0}{4} from NP~Ser.

The astrometric relationship of the X-ray source, the radio source,
and the visible object NP~Ser as understood prior to this work is
summarized by Deutsch et al. (1996).  Based on astrometric data used to
create the {\it HST} Guide Star Catalog (GSC), they concluded that the
long-accepted agreement between the positions of the radio
source and NP~Ser, the only real evidence implicating that star as the
counterpart, was likely spurious.  The extinction towards GX\,17+2, as
determined both from absorption in the X-ray spectrum and the observed
dust-scattered X-ray halo, is high ($A_V\,\squiggeqmm\,10$), and thus
a high-resolution search for a counterpart in the near infrared may be
more sensitive than previous work.  A ground-based IR search by Naylor
et al. (1991) did not yield any additional candidates besides NP Ser.
Here we discuss an improved astrometric alignment of optical and radio
frames, as well as new observations with NICMOS aboard {\it HST}.
A preliminary discussion of the NICMOS data has appeared in Wachter
(1998).

%==============================================================================
\section{OBSERVATIONS AND DATA REDUCTION}

\subsection{Archival Radio and X-ray Observations}

Grindlay \& Seaquist (1986) published a precise radio position for
GX\,17+2 based on 1982 VLA A array observations.  We have extracted and
recalibrated these and more recent observations from the VLA data archive,
and determined new positions and fluxes, using more modern positions
for the calibrators.  A summary of our results from the reprocessed
data is presented in Table 1.  Columns 1--3 list the observation date,
array configuration, and wavelength of the observation, respectively.
Column 4 lists the peak flux density or a $3\sigma$ upper limit and a
$1\sigma$ uncertainty.  The last two columns list $\alpha($B$1950)$
and $\delta($B$1950)$ and the respective $1\sigma$ uncertainties.
The early 1981 and 1982 positions agree well with the results of
Grindlay \& Seaquist (1986).  The later 1988 and 1989 position are also
consistent with each other, but disagree slightly in $\alpha$ with the earlier
observations.  As the later observations are viewed as more reliable than
very early VLA data, we determine our final best position by combining
the top three positions (1988--1989) in Table 1.  We then convert the
B1950 system coordinates into J2000 with the IDL {\it Astronomy User's
Library} (Landsman 1993) {\bf jprecess} procedure (calibrator epoch
1979.0) and give the final result, our preferred weighted radio position,
in Table 2.  This value differs slightly from the value in Deutsch et
al. (1996), which was not converted from the FK4 to FK5 system properly,
although the conclusions of that paper are not affected.  We note in
passing that the discrepancy between the 1981--1982 and 1988--1989
positions may include a component due to proper motion.

For completeness we have also extracted the three {\it Einstein} HRI X-ray
observations available in the HEASARC data archive, and recentroided the
positions of GX\,17+2.  These B1950 positions are combined into a single
best position, converted from B1950 to J2000 in the same fashion as above,
and also listed in Table 2.  As the precision is far less than the radio
data, the difference from Deutsch et al. (1996) is inconsequential.

\subsection{Optical Astrometry}

With the recent publication of the USNO-A2.0 star catalog (Monet
et al. 1998), it has become possible to tie an arbitrary field
rather easily to the International Celestial Reference Frame with
sub-arcsecond precision.  In a study of 283 extragalactic radio sources
which have very accurate VLBI radio positions and are well detected in
the USNO-A2.0 catalog, Deutsch (1999) reports that 90\% of the optical
positions are within \decsec{0}{40} of the radio coordinates and all
are within \decsec{0}{63}.  Also, when a solution is transferred from
the USNO-A2.0 catalog to modern epoch, deeper images (the POSS II in
Deutsch 1999) for 108 objects, 90\% of the optical positions are within
\decsec{0}{45} of the radio coordinates and all are within \decsec{0}{67};
this latter comparison may be the most relevant for the current situation.
Armed with this improved accuracy and quantification of the uncertainties
in the optical astrometric frame, we recompute an optical astrometric
solution for NP~Ser from the available imagery.

We select 84 USNO-A2.0 stars brighter than red magnitude 17 which
fall on an \aminbyamin{11}{11} ground-based R-band CCD image of this
field discussed by Deutsch et al. (1996).  Some of these stars are
then rejected due to nearby companions, image defects, or significant
proper motion.  The final set of 73 stars yields a solution with
residuals of $\sigma$=\decsec{0}{20} and therefore an approximate
uncertainty ($\sigma/\sqrt{n-3}$) in the alignment to the USNO-A2.0
frame of \decsec{0}{03}.  The frame is then transferred to the F675W-band
{\it HST} WFPC2 image presented by Deutsch et al. (1996) using 5 stars,
resulting in an alignment uncertainty of \decsec{0}{04}.

As there are only two well detected stars in common between the WFPC2
and NICMOS images, the astrometric scale and rotation in the NICMOS image
headers are preserved but the absolute offset is adjusted such that the
positions of these two stars agree to better than \decsec{0}{05} in the
NICMOS and WFPC2 images.  This process has yielded an alignment of the
NICMOS images to the USNO-A2.0 reference frame with an uncertainty of less
than \decsec{0}{1}.  Our final measurement of the optical position of NP
Ser on these various images in the USNO-A2.0 frame is listed in Table 2,
and is \decsec{0}{74} from the radio position, a discrepancy larger than
for {\it any} of the 108 solution transfer tests by Deutsch (1999).
The improved accuracy of the USNO-A2.0 frame now implies that this
current work supersedes the optical astrometry of Deutsch et al. (1996).

It must also be noted that NP~Ser itself does have an entry in the
USNO-A2.0 catalog as listed in Table 2.  This position is \decsec{0}{65}
from the radio position, a discrepancy which is larger than for {\it
any} of the 283 radio sources studied by Deutsch (1999), and thus the
association of NP Ser with the radio source is excluded at $>99$\%
confidence\,\footnote{\,Although the {\it HST} data suggest a slightly
larger offset of the optical and radio positions than the USNO-A2.0
data, the confidence of exclusion is slightly lower due to an additional
uncertainty term involving the transfer of frames and proper motions
(Deutsch 1999).}.  As we have measured deeper images obtained at
significantly better angular resolution, we are not alarmed by the
level of agreement of the two NP~Ser positions.  It is possible that
the discrepancy is a marginal detection of proper motion between the
1952.4-epoch USNO-A2.0 position and the 1995.8-epoch CCD frame.  Further,
NP Ser is among the fainter objects in the USNO-A2.0 catalog and thus
the centroiding uncertainty therein may be significant at this level.
For both reasons, we prefer our {\it HST} position, which is based on
the USNO-A2.0 positions of many reference stars, over the USNO-A2.0
catalog entry for NP Ser itself.  Most important, both positions imply
that NP Ser is not the radio or X-ray source.  Depending somewhat
on the catalog consulted, GX\,17+2 thus becomes the second or third
brightest persistent X-ray source on the celestial sphere lacking an
optical/infrared counterpart, a situation made yet more troubling given
the excellent radio position.

It should also be stressed that the radio and X-ray sources are known
to show correlated behavior (Penninx et al. 1988), and thus their
cross-identification seems nearly certain, quite aside from their
positional coincidence.  We need therefore not consider the region
inside the large X-ray error circle but well away from the precise radio
position.  For a relatively bright radio source such as this ($\sim1$~mJy
at 6~cm), the internal precision of the radio position determination
is very high, $<$\decsec{0}{1}. Future precise X-ray observations,
such as those from {\it Chandra X-ray Observatory}, will have internal
precision which, while excellent by X-ray astronomy standards, will not
equal this value, and will also in addition still be subject to similar
external astrometric uncertainties discussed here.  Thus while future
X-ray determinations will nicely complement this discussion, they are
unlikely to supersede it in accuracy.

\subsection{{\it HST} NICMOS Observations}

On 1997 August 14 we obtained brief {\it HST} NICMOS camera 1
(\decsec{0}{043} pixel$^{-1}$) images of NP Ser with the F110W and F160W
filters, which are similar to the Johnson J and H passbands, respectively.
A circular dither pattern of five exposures was used for each filter,
yielding a total exposure of 480~s in F110W and 1519~s in F160W.
These observations occurred early in NICMOS's lifetime, and many of the
reference files used in the original pipeline processing have since been
replaced with newer, in flight versions.  Thus we reprocess the raw data
with the newer reference files (current as of 1999 January) using the
STSDAS {\it calnica} task.  Further reduction is performed with software
written in IDL by E.~W.~D. or available in the IDL {\it Astronomy User's
Library} (Landsman 1993).

For each filter, the five exposures are mosaiced into a single frame
as follows.  First the fixed background pattern in each of the images
is removed by subtracting the median image derived from the stack
of images.  Next the bad pixels are zeroed in each individual exposure
and corresponding exposure mask.  The five images and masks are then
aligned using several reference stars.  Finally the images are summed
and the result divided by the sum of the exposure masks.

In Fig. 1 we show the mosaics of the GX\,17+2 field in F160W and F110W
as well as the WFPC2 F675W band.  The large circle indicates the $3''$
$\sim90$\% confidence radius error circle of the X-ray position
of GX\,17+2.  The small circle indicates the position of the radio
counterpart to the X-ray source.  The \decsec{0}{5} radius corresponds
to the 90\% confidence level in the alignment of the optical and radio
frames.  The bright star South of the radio circle is NP Ser itself.

\subsection{PSF Subtraction}

As NP Ser continues to be ruled out as the counterpart of GX\,17+2 with
good confidence on astrometric grounds, we examine the newly derived
optical/IR region corresponding to the variable radio source.  While the
bright central core of NP Ser is not within the radio error circle,
the complex wings of the {\it HST} NICMOS PSF visibly encroaches into
the region of interest.  We generate a synthetic PSF using version 4.4
of the Tiny TIM software package (Krist 1993) and attempt to subtract NP
Ser from our images.  The model PSF is slightly blurred with a Gaussian
filter to better match to the combined image, which suffers a small loss
of resolution in the mosaicing process and due to telescope jitter.
Figure 2 displays a \asecbyasec{4}{4} region centered on NP Ser in F160W
(top row) and F110W (bottom row) including the radio error circle from
Fig. 1.  The first panel in each row contains the original image; the
second panel shows the model PSF at approximately the same stretch;
the final panel displays the residual image after PSF subtraction.
The residuals are high within the central \decsec{0}{45} of NP Ser at
this stretch, but the subtraction is quite clean beyond that radius.

The PSF structure within the radio source error circle subtracts fairly
well, although note some residual of the four-point pattern remains.
In the F160W image, only two objects (indicated with arrows) remain
within the entire 90\% confidence error circle.  These objects, which
we denote A and B, are also clearly visible in the first panel before
the PSF subtraction.  We therefore suggest that only these two objects
are compatible with being the IR counterpart of GX\,17+2 to the depth
of all searches conducted thus far.  Our measured position for these
candidates are listed in Table 2.  The offsets of these objects from the
radio coordinates are only \decsec{0}{33} and \decsec{0}{39} for Stars
A and B, respectively, well within the 90\% systematic uncertainties in
the astrometric frame discussed in \S2.2.

The bottom row of Fig. 2 shows results of a similar PSF subtraction for
the F110W data, which is clearly not as deep.  Star A is clearly detected,
and there may be a $\sim2\sigma$ F110W detection of an object at the
location of Star B (indicated with an arrow), but this may instead be
a random fluctuation in the background.

Naylor et al. (1991) have reported from ground-based observations
infrared variability of the image of NP~Ser, and suggested that a
closely superposed counterpart of the X-ray source begins to contribute
significantly to the total IR flux from this region, thus explaining
their detection of modulation. As the new counterparts we discuss here are
cleanly separated from NP~Ser in our data, we can explicitly calculate the
ratio of $H$ fluxes of NP Ser to Stars A/B, which is $>10^2$.  Thus Naylor
et al. (1991) and Imamura et al. (1987) cannot have observed flux from
our candidates: they are simply far too faint.

\subsection{NICMOS Photometry}

Since the field is not very crowded by {\it HST} standards, we elect
to use aperture photometry to measure each of the objects detected in
this field.  A total of 133 objects are found, but 60 are not confidently
detected in the shallower F110W frame and only upper limits are available.
There are no objects detected in the F110W frame which are not detected
in the F160W frame.  Each object is measured with a 2 pixel radius
aperture and the result is adjusted with an aperture correction derived
from a suitable Tiny TIM PSF.  Count rates are converted to the VEGAMAG
system using the PHOTFNU keyword from the (reprocessed) headers and
$ZP(Vega)$ from column 3 of Table 18.1 in the {\it HST Data Handbook}.

The resulting magnitudes are similar to Johnson $J$ and $H$, but not
identical.  Since the final transformation depends on the stellar spectrum
and reddening in a complex fashion, we choose not to attempt to convert
our magnitudes to the Johnson system for all objects.  The VEGAMAG
magnitudes will be hereafter referred to as $m_{110}$ and $m_{160}$.
In some cases transformations between the Johnson and {\it HST} VEGAMAG
systems are calculated, based on the described assumptions.  We use the
STSDAS {\it synphot} package to calculate colors for all stars in Gunn
\& Stryker (1983) and interpolate the required transformation based on
the observed $(m_{110}-m_{160})$ color.  In the following discussion we
will quote $1\sigma$ relative uncertainties, although additional $\sim5$\%
systematic calibration uncertainties apply for each color.

%==============================================================================
\section{DISCUSSION}

\subsection{Color-Magnitude Diagram}

Figure 3 shows a color-magnitude diagram derived from aperture photometry
of the NICMOS F110W and F160W observations.  Plotted as black circles
are the 74 objects in the field which are detected in both frames.
Objects with only an $m_{110}$ upper limit are not plotted, but we denote
the $m_{110}\approx23.0$ upper limit as a dashed line.  Magnitudes are
in the VEGAMAG system, and formal $1\sigma$ uncertainties are provided
for each source.

The brightest object in the field is NP Ser at $m_{160}=14.75 \pm
0.02$, $(m_{110}-m_{160})=0.80 \pm 0.03$.  The open square indicates
the ground-based $H$, $(J-H)$ measurements of NP Ser reported by
Naylor et al. (1991), corrected to the {\it HST} VEGAMAG system, using
transformations $m_{110}=J+0.20$, $m_{160}=H+0.08$, appropriate for
a G star with $(m_{110}-m_{160})\approx0.8$.  The error bars on the
square denote the $1\sigma$ uncertainties quoted in Naylor et al. and
do not include an unknown additional uncertainty in the transformation.
Within the various uncertainties, our {\it HST} NICMOS measurement is in
agreement with the previous ground-based result.  Davidsen et al. (1976)
have spectroscopically typed the star as G, and Cowley, Hutchings \&
Crampton (1988) as G8, and clearly of dwarf or subgiant luminosity,
not giant.  We overplot on our color-magnitude diagram a stellar main
sequence with no reddening and a distance $d=1.5$ kpc for reference.
The distance is selected such that NP~Ser falls at late G on the sequence.

Two more dwarf stellar sequences are added to the diagram, one with
stars reddened by E$(B-V)=2.0$ at a distance $d=2.5$ kpc, and one with
stars reddened by E$(B-V)=3.0$ at a distance $d=4.0$ kpc.  We also add a
giant stellar sequence with stars reddened by E$(B-V)=4.4$ at a distance
$d=10$ kpc, which may best describe the reddest of the bright stars.
Clearly the color does not change drastically with stellar type, but
rather mostly with interstellar reddening.  This allows us to conclude
that except for NP~Ser itself, essentially all other stars in the field
must have E$(B-V)>2.0$.  In addition, at the fainter end there appears to
be a reasonably narrow locus of stars corresponding to E$(B-V)\sim3.0$.
These conclusions do not depend on the assumption that the field stars
are actually of dwarf luminosity, and some may be giants, although the
reddening per unit distance would then be rather low.

\subsection{Nature of the Infrared Candidates}

As pointed out in \S2.4, there are now only two viable candidates based
on the latest optical astrometric alignment.  For Star A we measure
$m_{160}=19.93 \pm 0.02$, $(m_{110}-m_{160})=2.08 \pm 0.09$ after
subtracting the NP~Ser PSF.  For Star B, we find $m_{160}=20.94 \pm
0.04$, and there may be be a faint detection of this object in the F110W
image, although alternatively it may be a PSF residual or image defect.
We report the possible detection with $m_{110}\,\squiggeqmm\,23.1$,
and label the object with a triangle in Fig. 3.  For objects with color
$(m_{110}-m_{160})\approx2.0$, transformations to Johnson magnitudes
of $(J-m_{110})\approx-0.5$ and $(H-m_{160})\approx-0.10$ are implied.
Approximate magnitudes for Star~A of $H\approx19.8$, $(J-H)\approx1.7$,
and for Star B or $H\approx20.8$, $(J-H)\,\squiggeqmm\,1.8$ may thus
be inferred.

Of course, one or perhaps both of these candidates may yet be
further chance superpositions!  In our \asecbyasec{13}{14} F160W-band
field of view, we detect 0.7 objects arcsec$^{-2}$ with $m_{160}<22.2$.
As our 90\% confidence radio error circle has area 0.8~arcsec$^{2}$, it
is easily inferred that the likelihood of one of these unrelated objects
falling within our error circle is close to 50\%.  As with most {\it a
posteriori} probability calculations, this one is extremely sensitive to
arbitrary assumptions.  Had we considered a $1\sigma$ confidence radio
error circle, for example, the chance coincidence probability would
almost halve.

Obviously a search for temporal variability of these candidates, and
particularly spectroscopy of these stars, are future steps forward in the
problem.  Even for the brighter of the two, these will be challenging
observations due to the small angular separation, \decsec{0}{9},
and large magnitude difference, $\Delta m_{160}=5.2$, from NP Ser,
as well as the small separation between A and B (\decsec{0}{5}).
However, our photometry, combined with extinction estimates here and
made previously in the literature, already allow some simple but quite
interesting inferences.

Previous constraints on the extinction to GX\,17+2 from X-ray observations
have been summarized by Naylor et al. (1991), and go beyond the
normal procedure of fitting the turnover in the soft X-ray spectrum,
as independent data from a dust scattered X-ray halo (Catura 1983) are
also available.  These estimates are consistent, and yield $11 < A_V <
14$. From our color-magnitude diagram (Fig.~3) we have concluded above
that all objects in the field (except the presumptively irrelevant NP~Ser)
probably have $A_V > 6$, and many have $A_V > 9$, yet another consistent,
albeit uncertain, estimate.  We adopt $A_V = 5\,A_H$ (Cardelli, Clayton,
\& Mathis 1989) and $A_H=2$ for GX\,17+2 for the remainder of this
discussion;  our most important conclusions are not unduly sensitive to
this assumption.

Although the distance to GX\,17+2 is poorly constrained, its projected
location near the galactic center, the inferred extinction, as well
as the observed X-ray flux and spectrum (Christian \& Swank 1997) and
characteristics of the X-ray bursts (Sztajno et al. 1986) all suggest that
$5 < d < 20$~kpc is almost surely a very conservative range that includes
the actual distance of the object. Including a correction for extinction,
this implies $4.5 > M_H > 1.5$ for the brighter of the two candidates.

$H$-band photometric data on Z-sources, for comparison with the results
we obtain for GX\,17+2, are relatively sparse in the literature.
The two best studied sources are probably, not surprisingly, Sco~X--1
and Cyg~X--2, although their distances, and thus $M_H$ values, are
still only approximately known.  Luckily their extinction at $H$ is
probably small.  The Z-source GX\,349+2, although more poorly studied,
is confidently optically identified.  The current best estimates of $H$,
reddening, distance, and the inferred $M_H$ for these three sources are
collected from the literature in Table 1.  As these objects are
all photometrically variable, these values must be regarded as typical
rather than precise. We see values in the range $-1.3 < M_H < -0.3$
are represented.

We are therefore faced with the surprise that, despite the very similar
X-ray behavior of GX\,17+2 to that of the other Z-sources, at least at
the time of our observations the optical/IR counterpart is intrinsically
fainter, quite possibly by $\sim10^2$, than previously studied analogs.
Furthermore, if Star A is not the true counterpart of the X-ray source,
this conclusion is only strengthened, as there is nothing brighter within
the radio error circle.

As at least the visible light from low mass X-ray binaries is typically
dominated by reprocessing of the intense incident X-ray flux, we have
considered if by bad luck the X-ray source could have been quiescent
during our {\it HST} observations, although there is no incident in
the literature of such dramatic behavior. The public data archive of
the All Sky Monitor of the {\it Rossi X-ray Timing Explorer} allows us
to effectively rule out this possibility.  Although no {\it RXTE} ASM
measurements coincide precisely with the NICMOS observations, a dozen
measurements on the same day indicate modest variability about the mean
intensity $\sim700~\mu$Jy, the same intensity value tabulated in the
compendium of van Paradijs (1995). We thus believe that this unusually
low luminosity of the counterpart is a genuine feature of the source. An
obvious but interesting caveat is that if the system inclination is
favorable and the timing of our observations highly fortuitous, we could
have observed the system in a brief eclipse which has thus far escaped
detection in X-rays. Even a small number of further infrared
observation should clarify this possibility.

Barring this unlikely eclipse scenario, the inferred $L_x/L_{opt}$
for this source is extremely high (a longstanding problem even if
NP Ser were the counterpart).  Bradt \& McClintock (1983) derived
$L_x/L_{opt}\sim3000$ based on the assumption that NP~Ser is the
identification with E$(B-V)=0.6$; this $L_x/L_{opt}$ is higher than
any other LMXB in their compilation.  More current determinations of
E$(B-V)$ for NP~Ser find it to be almost negligible, which implies
an even higher value of $L_x/L_{opt}$.  If instead the true optical
counterpart lies within the radio error circle we have searched (Fig. 2),
we can estimate a lower bound to $L_x/L_{opt}$ as follows: by assuming
$F_X>700\ \mu$Jy, an intrinsic spectrum of a 25,000 K blackbody, $A_H<3.0$
(based on X-ray constraints), and $H>19.8$ (Star A or fainter)
we find $L_x/L_{opt}>1800$.  This is $3\times$ higher than the other
Z-sources and the LMXB average (van Paradijs \& McClintock 1995),
but comparable to the remarkable 11~min binary X1820--303,
which may exhibit the highest value of $L_x/L_{opt}$.  We infer that
the optical/IR candidate of GX\,17+2 should be highly variable,
yet another piece of evidence against NP~Ser.

\section{CONCLUSIONS}

A new astrometric solution referring optical and infrared objects
in the field of GX\,17+2 to the International Celestial Reference
Frame eliminates with high confidence NP~Ser as the counterpart of the
non-thermal, variable radio source. If we also consider the co-variability
of the X-ray and radio objects (Penninx et al. 1988), NP~Ser cannot
be the counterpart of the X-ray source, a conclusion which has been
mooted on and off over three decades, but never entirely accepted. New
{\it HST} NICMOS observations identify two faint ($H\approx19.8, 20.8$)
stellar objects coincident with the radio position, although they may
be additional chance superpositions of unrelated objects. Regardless of
this possibility, the evidence is strong that the optical/IR counterpart
of GX\,17+2 is surprisingly underluminous compared with other Z-sources.
We expect marked photometric variability from the counterpart.

Even after thirty years of observations, GX\,17+2, though one of the most
intense X-ray sources known, continues in many ways to defy a simple
analysis.

%==============================================================================
\acknowledgments

We thank Eddie Bergeron for his assistance with reprocessing and
photometric calibration of the NICMOS data, Mike Shara for the R-band
CCD frame, and Bob Hjellming for discussions about his 1981 and 1989
data.  Support for this work was provided by NASA through grant number
GO-073366.01-96A from the STScI, which is operated by AURA, Inc.

%==============================================================================

\clearpage

\begin{deluxetable}{lcrrcc}
\tablenum{1}
\tablecolumns{6}
\tablecaption{VLA Flux and Position Measurements\,\tablenotemark{a}}
\tablehead{
\colhead{Observation} &
\colhead{Array} &
\colhead{$\lambda$} &
\colhead{Flux Density} &
\colhead{} &
\colhead{} \\
\colhead{Date} &
\colhead{Config} &
\colhead{(cm)} &
\colhead{(mJy)} &
\colhead{$\alpha($B$1950)$} &
\colhead{$\delta($B$1950)$} 
}
\startdata 
1989 Mar 31 & B &3.6& $ 0.87\ (0.04)$  & 18 13 10.922 (0.002) & $-14$ 03 14.47 (0.03) \nl
1989 Mar 31 & B & 6 & $ 0.90\ (0.05)$  & 18 13 10.919 (0.004) & $-14$ 03 14.45 (0.06) \nl
1989 Mar 31 & B &20 & $<0.2\ \ (0.06)$  & & \nl
1988 Apr 05 & C & 2 & $ 3.1\ \ (0.1)\ \,$ & 18 13 10.921 (0.003) & $-14$ 03 14.42 (0.05) \nl
1982 Jun 21\,\tablenotemark{b} & A & 6 & $<0.3\ \ (0.1)\ \,$ & & \nl
1982 Feb 21\,\tablenotemark{b} & A & 6 & $ 1.5\ \ (0.2)\ \,$ & 18 13 10.931 (0.004) & $-14$ 03 14.52 (0.07) \nl
1981 Mar 19 & A & 6 & $ 1.19\ (0.07)$  & 18 13 10.929 (0.003) & $-14$ 03 14.47 (0.05) \nl
\enddata
\tablenotetext{a}{\,Weighted, preferred J2000 position is given in Table 2.}
\tablenotetext{b}{\,Rereduction with modern calibration of the data reported by Grindlay \& Seaquist (1986).}
\end{deluxetable}

% -----------------------------------------------------------------------------
%\clearpage
{\small

\begin{deluxetable}{llllll|rrr}
\tablenum{2}
\tablecolumns{9}
\tablecaption{J2000 Positions in the GX 17+2 field}
\tablehead{
\colhead{} &
\colhead{} &
\colhead{} &
\colhead{} &
\colhead{$\sigma_\alpha$} &
\colhead{$\sigma_\delta$} &
\colhead{$\Delta\alpha$\,\tablenotemark{a}} &
\colhead{$\Delta\delta$\,\tablenotemark{a}} &
\colhead{$\Delta$R\,\tablenotemark{a}} \\
\colhead{Object} &
\colhead{Observation} &
\colhead{$\alpha($J$2000)$} &
\colhead{$\delta($J$2000)$} &
\colhead{$('')$} &
\colhead{$('')$} &
\colhead{$('')$} &
\colhead{$('')$} &
\colhead{$('')$}
}
\startdata 
GX 17+2 & VLA 2-6 cm           & 18 16  1.389 & $-14$ 02 10.62  & 0.06  &  0.04 & ....... & ....... & ....... \nl
GX 17+2 & {\it Einstein} X-ray & 18 16  1.25  & $-14$ 02 10.9   & 2.8   &  2.8  & $-2.0\ \,$ & $-0.3\ \,$ &  $2.0\ \,$ \nl
NP Ser  & USNO-A2.0            & 18 16  1.356 & $-14$ 02 11.06  & 0.18  &  0.18 & $-0.48$ & $-0.44$ &  0.65 \nl
NP Ser  & R-band CCD           & 18 16  1.380 & $-14$ 02 11.34  & 0.25  &  0.25 & $-0.13$ & $-0.72$ &  0.73 \nl
NP Ser  & WFPC2 F675W          & 18 16  1.373 & $-14$ 02 11.32  & 0.25  &  0.25 & $-0.23$ & $-0.70$ &  0.74 \nl
NP Ser  & NICMOS F160W         & 18 16  1.373 & $-14$ 02 11.32  & 0.25  &  0.25 & $-0.23$ & $-0.70$ &  0.74 \nl
Star A  & NICMOS F160W         & 18 16  1.373 & $-14$ 02 10.38  & 0.25  &  0.25 & $-0.23$ & $ 0.24$ &  0.33 \nl
Star B  & NICMOS F160W         & 18 16  1.406 & $-14$ 02 10.32  & 0.25  &  0.25 & $ 0.25$ & $ 0.30$ &  0.39 \nl
\enddata
\tablenotetext{a}{\,offset from weighted, preferred radio position}
\end{deluxetable}

}

% -----------------------------------------------------------------------------
%\clearpage

\begin{deluxetable}{lccccl}
\tablenum{3}
\tablecolumns{6}
\tablecaption{Photometric Properties of Z Sources}
\tablehead{
\colhead{Object} &
\colhead{$H$} &
\colhead{E($B-V$)} &
\colhead{d (kpc)} &
\colhead{$M_H$} &
\colhead{Ref.} 
}
\startdata 
Sco~X--1  & 11.3                   &  0.3   &  2.0 &   $-0.3$   & 1,2,3 \nl
GX~349+2  & 14.7                   &  1.4   &  9.0 &   $-0.8$   & 1,4,5 \nl
Cyg~X--2  & 13.4\tablenotemark{a}  &  0.4   &  8.0 &   $-1.3$   & 1,6,7,8 \nl
\enddata
\tablenotetext{a}{The mass donor probably contributes a substantial fraction of 
the IR brightness}
\tablerefs{1.~Wachter (1998), 2.~Bandyopadhyay et al. (1999), 3. Bradshaw 
et al. (1997), 4.~Wachter \& Hoard (1999) 5.~Penninx (1989),
6.~McClintock et al. (1984), 7.~Orosz \& Kuulkers (1999), 8.~Cowley et al.
(1979)}
\end{deluxetable}

%\clearpage

\begin{figure}
\plotone{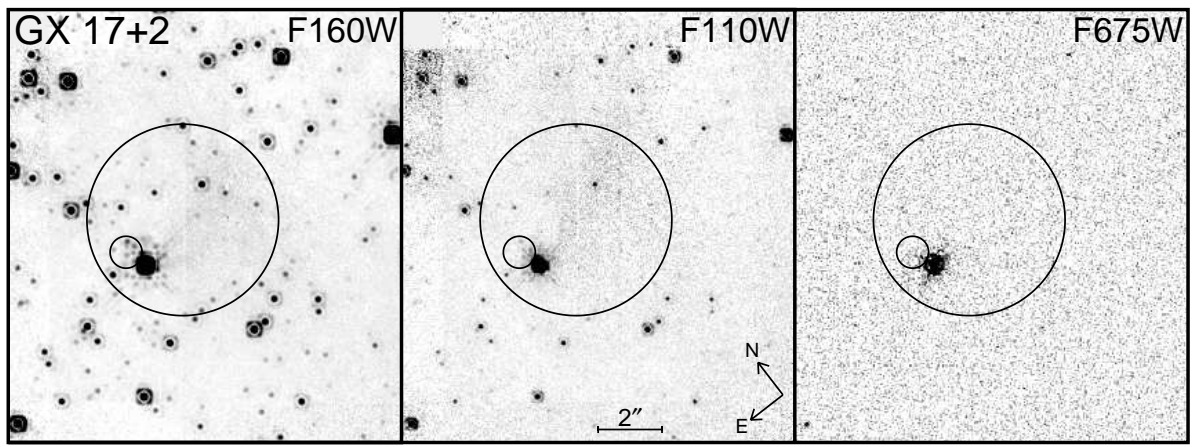}
\caption{Approximately \asecbyasec{13}{14} H-, J-, and R-band images
of the field of GX\,17+2, obtained with {\it HST} NICMOS and WFPC2.
The large circle indicates the $3''$ $\sim90$\% confidence radius error
circle of the X-ray position of GX\,17+2.  The small circle indicates the
position of the associated radio source, as derived from our optical/IR
astrometric solution.  The \decsec{0}{5} radius corresponds to the 90\%
confidence level in the alignment of the optical and radio frames.
The bright star South of the radio circle is NP Ser.  Note the slight
cardinal rotation of the frames.}
\end{figure}

\begin{figure}
\plotone{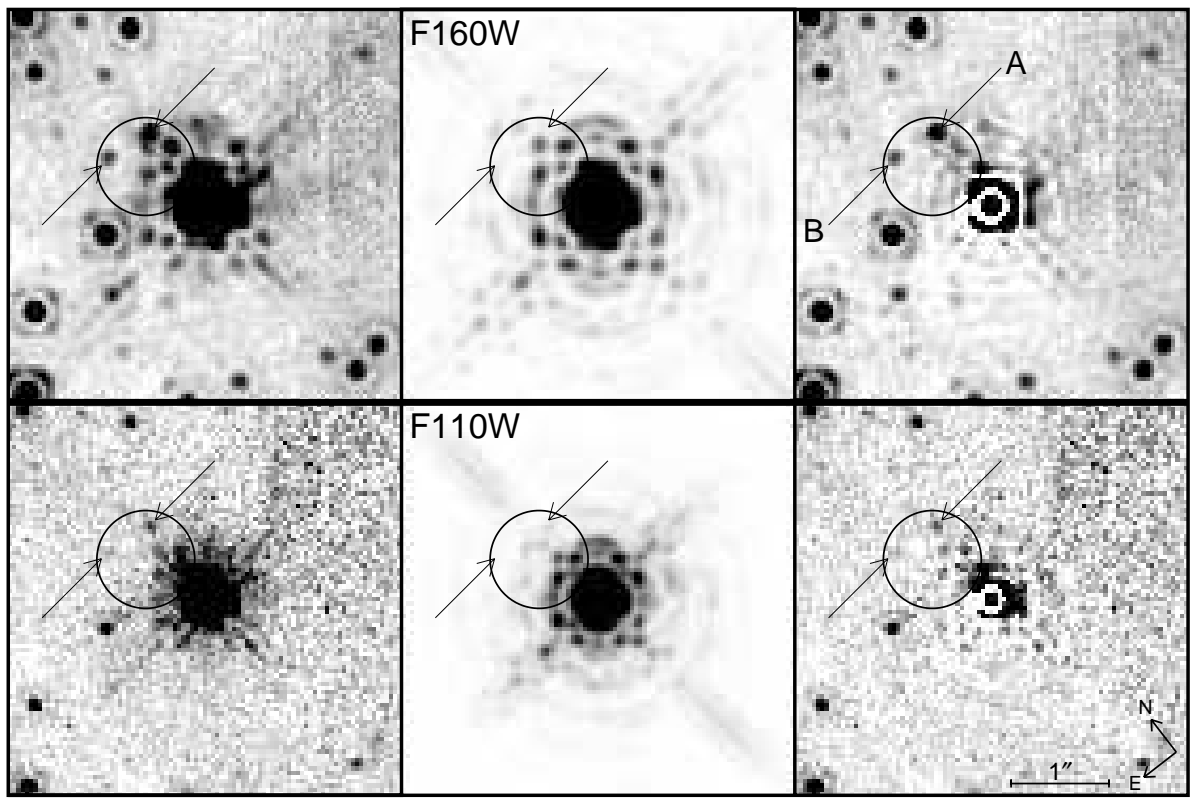}
\caption{\asecbyasec{4}{4} {\it HST} NICMOS images surrounding the
bright star NP Ser.  The 90\% confidence \decsec{0}{5} radius circle
indicates the position of the radio source.  The top row displays the
F160W (H-band) images, and the bottom row the F110W (J-band) images.
{\it Left:} the original summed image; {\it center:} the model PSF
at approximately the same stretch; {\it right:} the residual image
after PSF subtraction.  Note that the large majority of flux within the
radio error circle is due to the bright PSF wings of NP~Ser.  Only two
objects {\it(arrows)} remain within the entire 90\% confidence error
circle after PSF subtraction in the F160W image.  The objects are also
clearly visible in the first panel before the PSF subtraction.  Star A
is also detected in the F110W image, and there may be a $\sim2\sigma$
F110W detection of an object at the location of Star B {\it(arrow)},
but this may well instead be a random fluctuation in the background.}
\end{figure}

\begin{figure}
\plotone{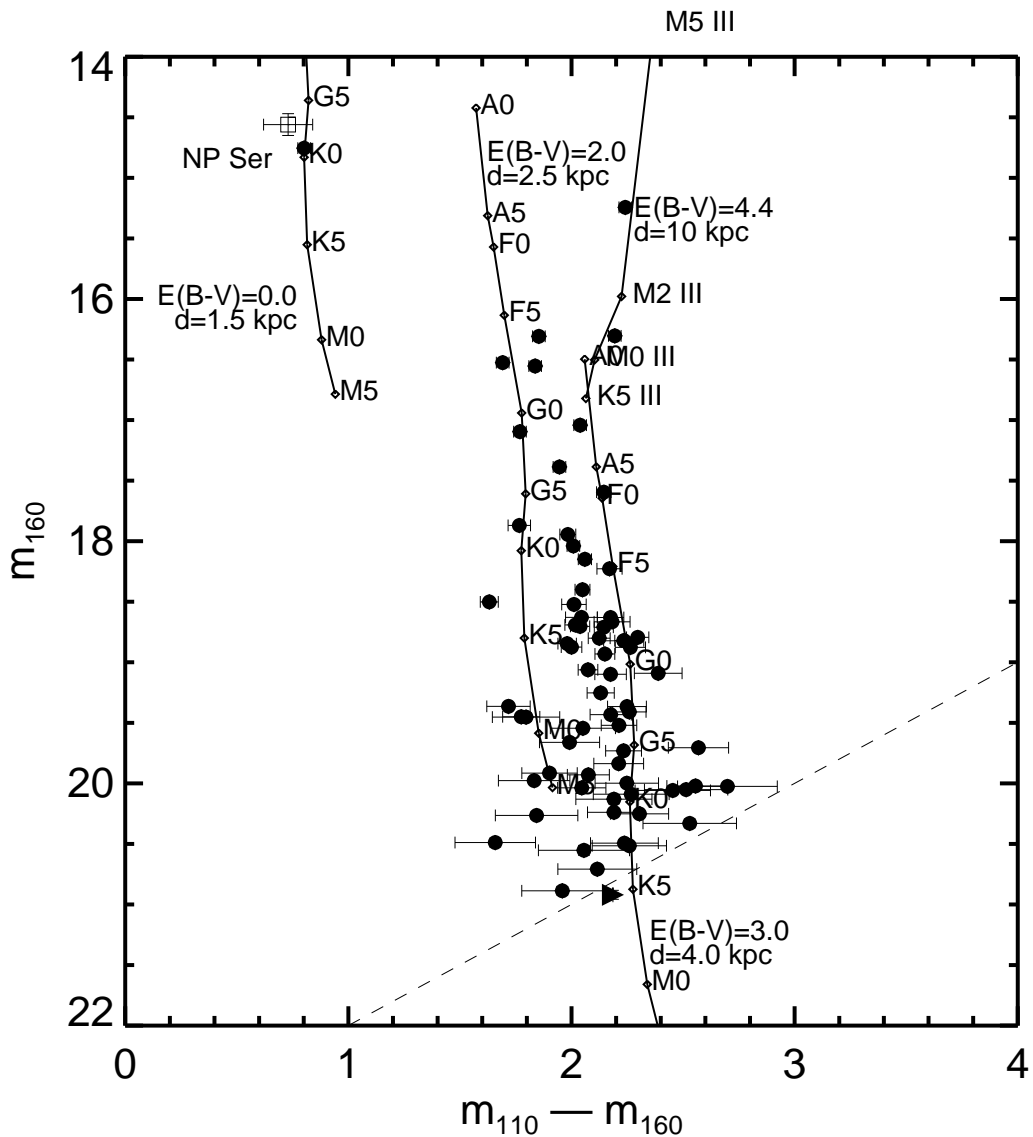}
\caption{{\it HST} NICMOS color-magnitude diagram for the
\asecbyasec{13}{14} region surrounding GX\,17+2.  {\it Open square:}
previously published data for NP Ser, now thought to be a chance
superposition on the radio/X-ray source.  The solid circle nearby is our
{\it HST} measurement; {\it vertical lines:} loci of main sequence and
giant stars for various assumed reddening and distance; {\it dashed line:}
bound implied by the approximate $m_{110}\approx23.0$ detection limit in
our data.  These data indicate that while NP Ser is likely a foreground
object, other detected objects in the field have E$(B-V)\,\squiggeqmm\,2$,
and many may have E$(B-V)\,\squiggeqmm\,3$.  One candidate for the IR
counterpart of GX\,17+2 appears near the bottom as a filled triangle.
Although it is very well detected at $m_{160}$, it is only marginally
seen at $m_{110}$, and thus is designated as a lower limit in color.}
\end{figure}

\end{document}